\documentclass[twocolumn,aip,apl,floats]{revtex4}
\usepackage{epsfig}
\usepackage{dcolumn}
\usepackage{bm}
\usepackage[latin1]{inputenc}
\usepackage{amsmath,amsfonts,amssymb,amsthm,mathrsfs}
\DeclareTextSymbol{\degre}{OT1}{23}

\begin{document}
\title{Bulk Cr tips for scanning tunneling microscopy and spin-polarized scanning tunneling microscopy}
\author{A. Li Bassi, C. S. Casari, D. Cattaneo, F. Donati, S. Foglio, M. Passoni, C. E. Bottani}
\address{NEMAS - Center for NanoEngineered MAterials and Surfaces
\\ \mbox{CNISM - Dipartimento di Ingegneria Nucleare, Politecnico di
Milano}\\ Via Ponzio 34/3, I-20133, Milano, Italy}
\author{P. Biagioni, A. Brambilla, M. Finazzi, F. Ciccacci, L. Duò}
\address{NEMAS - Center for NanoEngineered MAterials and Surfaces
\\ CNISM - Dipartimento di Fisica, Politecnico di Milano\\ Piazza
Leonardo da Vinci 32, I-20133, Milano, Italy}
%\date{\today}

\begin{abstract}
A simple, reliable method for preparation of bulk Cr tips for
Scanning Tunneling Microscopy (STM) is proposed and its
potentialities in performing high-quality and high-resolution STM
and Spin Polarized-STM (SP-STM) are investigated. Cr tips show
atomic resolution on ordered surfaces. Contrary to what happens
with conventional W tips, rest atoms of the Si(111)-7$\times$7
reconstruction can be routinely observed, probably due to a
different electronic structure of the tip apex. SP-STM
measurements of the Cr(001) surface showing magnetic contrast are
reported. Our results reveal that the peculiar properties of these
tips can be suited in a number of STM experimental situations.
\end{abstract}

\maketitle

Scanning tunneling microscopy (STM) is widely exploited to study
surfaces with atomic resolution. Starting from its invention, it
has been performed using tunneling tips fabricated with a great
variety of materials. So far, the most commonly adopted tips are
prepared by electrochemical etching of W wires \cite{melmed}. For
some specific applications, however, other materials are required.
For example, in Spin Polarized STM (SP-STM) measurements,
ferromagnetic or anti-ferromagnetic tips exhibiting an intrinsic
spin polarization have to be used. Usual ferromagnetic tips are
made of Fe\cite{koltunn}, Ni \cite{cavallini}, Co \cite{albonetti}
or Fe coated W tips \cite{bode}, while anti-ferromagnetic tips
have been prepared using MnNi \cite{murphy1}, MnPt \cite{murphy2},
Cr \cite{wiesen91,shvets, ceballos}, Cr-coated \cite{kubetzka} or
Mn-coated \cite{yang} W tips . Anti-ferromagnetic materials are
usually preferred because they do not exhibit significant
perturbing stray field and are not influenced by external fields.
Cr is the only metal with a (bulk) N$\acute{e}$el temperature (311
K) above room temperature (RT) and this makes it interesting in
non cryogenic SP-STM. MnNi and MnPt alloys have a higher
N$\acute{e}$el temperature but, usually, tips obtained by etching
are characterized by a lack of stoichiometry that could produce a
net magnetization of the apex. These considerations motivate the
aim of this work, which consists in the development of a simple
preparation method of bulk Cr tips and in a first investigation of
their properties as STM and SP-STM probes.

Previous results concerning the preparation and use of bulk Cr
tips for STM are reported in Refs.
\onlinecite{wiesen91,shvets,ceballos}. In particular, in Refs.
\onlinecite{wiesen91,shvets} Cr tips have been prepared by etching
of Cr rods in KOH solution and subsequent mechanical breaking in
ultra-high vacuum (UHV). More recently, they have been obtained by
electrochemical etching of cylindrical rods in NaOH solution
\cite{ceballos}. Both these methods lead to tips with a good
aspect ratio. Only in Refs. \onlinecite{wiesen91,shvets} it has
been shown that Cr tips can achieve atomic resolution on the
Si(111)-7$\times$7 reconstructed surface, while SP-STM
measurements are not reported. It must be observed that both
procedures contain some delicate and complex steps: either
breaking an etched rod in UHV in one case, or shaping the rod to
cylindrical section in the other. These aspects could make the
realization of Cr tips rather complicated. We developed a
simplified procedure which avoids the above mentioned steps. We
started from rods of polycrystalline Cr with a nearly square cross
section of 0.7 mm $\times$ 0.7 mm obtained by cutting a 99.99 \%
Cr foil. Asymmetrical shape of the rods due to lack of uniformity
in the foil thickness can often result in a bad aspect ratio,
nevertheless atomic resolution on atomically flat surfaces could
be easily achieved anyway, as described below. The etching
procedure can be divided in two steps; in the first, we perform a
pre-etching, with a ring-shaped gold cathode, applying a DC
voltage in the 5-7 V range in order to reduce the rod
cross-section, while in the second step etching is performed using
a DC voltage in the 3-4 V range. Both NaOH and KOH 1.5 M solutions
have been tested with good results. During the etching we observe
formation of Na/K and Cr compounds on the surface of the rod.
These compounds are soluble in water and can be removed by
stopping the pre-etching and washing the rod in a water ultrasonic
bath. A lower voltage is used in the second step to limit the
accumulation of the compounds on the rod. SEM images of the
various Cr tips obtained after the described procedure have been
acquired (not shown). In general, we observed that even though the
overall shape on a micron scale is usually far from being regular,
the tip apex is sharp at least at the observed scale (tens nm).

\begin{figure}
 \includegraphics[scale=0.6]{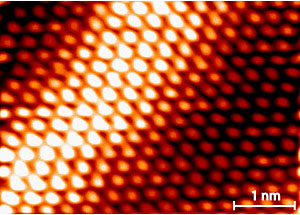}\
  \caption{Image of Au(111) surface taken at V$_b$=1 V, I$_b$=1 nA and at T=300 K, showing superstructure and atomic resolution.}\label{fig1}
\end{figure}

RT, constant current STM measurements with Cr tips have been
performed using a Omicron UHV VT-SPM. The tips were tested on the
Au(111) surface, as a typical example of a metallic system. In
Fig. \ref{fig1} a topographic image of this surface is shown. It
can be observed that atomic resolution, clear separation among
atoms as well as the characteristic superstructure are easily
achieved.

\begin{figure}
  \includegraphics[scale=0.55]{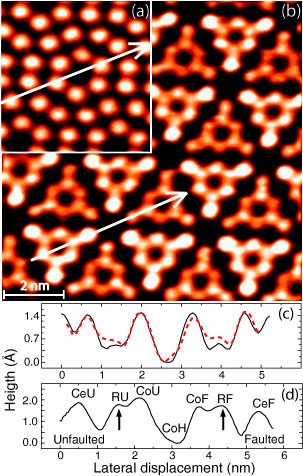}
  \caption {Images of Si(111)-7$\times$7 surface taken at (a)
  V$_b$=1 V and I$_b$=1 nA and (b) V$_b$=-1.5 V and I$_b$=2.2 nA.
  The presence of rest atoms (R) can be observed between the
  corner (Co) and the two center adatoms (Ce), in both faulted (F) and unfaulted (U) half (CoH indicates the corner hole of the cell).
  (c) Line profile along the diagonal of the cell of case (a), taken with a Cr tip (black solid line), vs. normalized line profile taken with a W tip at V$_b$=1 V (red dashed line).
  (d) Line profile along the diagonal of the cell of case (b).} \label{fig2}
\end{figure}

The Si(111)-7$\times$7 surface was chosen as the main test
surface. It is a very well known surface, both theoretically and
experimentally. It is also one of the most investigated surfaces
with STM (see e.g. Ref. \onlinecite{wiesen94} and references
therein). The complex structure of its 7$\times$7 unit cell
\cite{taka} makes it ideal in order to test the capability of
resolving detailed energetic and spatial features with sub-nm
resolution. It is well known that, with commonly used tips, at
positive applied bias the topographic STM image clearly shows the
twelve adatoms \cite{binnig}, while at negative bias the
corrugation amplitude is worse. In usual conditions rest atoms,
which lie about 0.7 Å below the nearest corner adatom, are not
observed: they have been detected using particular semiconducting
tips, able to suppress the signal from the adatoms \cite{sutter}.
In a recent paper \cite{wang}, the rest atoms have been imaged
using W tips with exceptionally high aspect ratio: they have been
resolved only at the bias of -1.5 V and a net distinction from the
corner adatoms is evident only in the unfaulted half. In Fig.
\ref{fig2}(a) we report a representative image of the
Si(111)-7$\times$7 surface at +1 V bias obtained using a Cr bulk
tip, where excellent separation between adatoms can be
appreciated. In Fig. \ref{fig2}(b) the same surface is then shown
at -1.5 V. One can clearly observe the distinction between faulted
and unfaulted half cell, the net separation between different
cells and, above all, the presence of all the rest atoms belonging
to the unit cell. Line profiles, like the one shown in Fig.
\ref{fig2}(d), allow distinguishing rest atoms both in the
unfaulted and faulted half of the cell. While Wang et al.
\cite{wang} claimed that rest atoms could be observed only with
two of the many W tips used, here, using bulk Cr tips, we observed
rest atoms routinely, with all the tips we used. This is a first
important evidence that Cr bulk tips allow for a peculiar
sensitivity, reducing convolution effects of the tip and enhancing
its capability of resolving details in the observed topographic
and electronic structure.

A different tip sensitivity can be ascribed both/either to its
geometrical configuration (local curvature radius, aspect ratio)
and/or to its local electronic density of states (LDOS) properties
and to the spatial distribution of the apex atom electronic
orbitals. In the work by Wang et al. a simple model of the surface
line profile as a function of the tip curvature radius is
discussed, observing that it is necessary to use a tip with an
equivalent radius of about 7 Å in order to distinguish at least
the rest atom of the faulted half. Since our result was obtained
for all the Cr tips used, it may be not sufficient to explain it
only in terms of a very sharp geometrical shape of the apex,
because such a control and reproducibility of the final tip
curvature radius and aspect ratio appears beyond the possibilities
of our simplified fabrication method. Therefore, electronic
properties and in particular the local density of states at the Cr
tip last atom, which is reasonably nearly the same in all the
tips, could be the factor that mostly contributes to the peculiar
properties observed.

As a further investigation, we compared performances of Cr bulk
tips with home-made W tips obtained with a standard etching
procedure \cite{melmed}. Line profiles of the same Si
(111)-7$\times$7 surface obtained with Cr and W tips, both at
$V_b$=1 V and $I_b$=0.5 nA, are compared in Fig. \ref{fig2}(c). We
evaluated the line profiles of these images along a cell diagonal
and we normalized them so that the maximum height excursion
(corresponding to the excursion between the corner adatom and the
corner hole) is the same for the W and the Cr tip. By doing so we
want to compare the imaging contrast and resolution of the tips,
normalizing for effects related to different barrier heights and
different tip-to-sample distance. It can be noted that there are
no significant differences both in the height contrast and in the
lateral atomic resolution. This again supports the conclusion that
the observation of rest atoms at -1.5 V bias is probably related
to the Cr apex electronic structure rather than to a geometrical
factor.

\begin{figure}
  \includegraphics[scale=0.55]{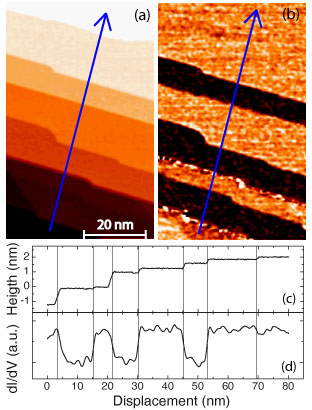}
  \caption{(a) Topographic and (b) differential conductivity maps
   of Cr (001) surface taken at V$_b$=-0.37 V and I$_b$=1 nA.
   Line profiles (c) and (d) clearly show the magnetic contrast obtained with Cr bulk tip. The image contrast of the differential
   conductivity map is $\sim$ 15 \% while the signal-to-noise ratio is $\sim$ 4.}\label{fig3}
\end{figure}

All these results provide strong arguments towards the interest in
the performances of bulk Cr STM tips. If combined with the
anti-ferromagnetic properties of Cr, SP-STM and scanning tunneling
spectroscopy (STS) experiments exploiting the advantages of an
elemental bulk tip with enhanced sensitivity are foreseen. We
performed preliminary SP-STM measurements with these tips choosing
one of the most investigated magnetic surfaces via SP-STM/STS
techniques, namely the Cr(001) surface. It is well known that in
this system terraces with opposite magnetic in-plane polarization
can exist \cite{blugel} and can be directly observed performing
SP-STM/STS measurements \cite{kleiber}. In Fig. \ref{fig3} we show
a topographic image taken at $V_b$=$-0.37$ V (a) and a
differential conductivity map (b) acquired at the same bias with
our bulk Cr tip. The corresponding line profiles are also shown.
Even if a direct quantitative comparison with previously reported
data on clean Cr(001) is not possibile because of the presence of
contaminants on our sample \cite{schmid}, nevertheless the
capability of resolving the magnetic contrast is clearly evident.
Therefore, bulk Cr tips exhibit in-plane magnetic sensitivity,
leading to the possibility of achieving SP-STM with them.

In conclusion, we have presented a simplified method to produce Cr
bulk tips and an investigation of their potentiality as STM
probes. Lateral sensitivity of these tips, probably due to the
features of the tip LDOS structure, can significantly enhance the
quality of the STM image of an atomically flat surface. Atomic
resolution on Si(111)7$\times$7 has been easily achieved and rest
atoms of the Si (111)-7$\times$7 reconstruction have been
routinely resolved. Successful tests on the magnetic Cr(001)
surface have been conducted, demonstrating the feasibility of
performing SP-STM measurements. We believe that this work opens
the way for a useful introduction of this kind of STM probes in
(SP-)STM/STS experiments, which allows avoiding either in-situ
evaporation or use of non elemental materials for tip
preparations.

We thank A. Cricenti and M. Capozi for inspiring discussions. This
work was carried out in the framework of Progetto Innesco CNISM
2006.

%\newpage

%\newpage

\end{document}